\documentclass[reprint,superscriptaddress,aps,pre]{revtex4-2}
\usepackage[utf8]{inputenc}

\usepackage{amssymb}
\usepackage{amsmath}
\usepackage{graphicx}
\usepackage{colortbl}
\usepackage[dvipsnames]{xcolor}
\usepackage[caption=false]{subfig}
\usepackage[hidelinks]{hyperref}
\usepackage{comment}
\usepackage[english]{babel}

\newcommand{\dif}{\text{d}}
\newcommand{\Dif}{\mathcal{D}}
\newcommand{\mean}[1]{\left\langle #1 \right\rangle}
\newcommand{\MBR}{\mathcal{M}}

\newcommand{\T}{\intercal}

\newcommand{\cut}{l}
\newcommand{\cutq}{l}

\newcommand{\reg}{\vartheta_\cut}
\newcommand{\regq}{\vartheta_\cutq}

\date{August 2023}

\begin{abstract}
    Correlation functions are a standard tool for analyzing statistical particle trajectories. Recently, a so called mean back relaxation (MBR) has been introduced, which correlates positions at three time points. The deviation of its long time value from $\frac{1}{2}$ has been shown to be a marker for breakage of time reversal symmetry for confined particles. Here, we extend the analysis of MBR in several ways, including discussion of a cut off length used when evaluating MBR from trajectory data.  Using a path integral approach, we provide a general expression for MBR in terms of multipoint density correlations. For Gaussian systems, this expression yields a relation between MBR and mean squared displacement.
    We finally demonstrate that MBR can be applied to other stochastic observables besides particle position. Using it for microscopic densities, its deviation from $\frac{1}{2}$ is a marker for broken detailed balance in confinement or in bulk systems. 
    
\end{abstract}

\begin{document}

%\title{Properties of conditioned correlations}
\title{Mean Back Relaxation for Position and Densities}
\author{Gabriel Knotz}
\address{Institute for Theoretical Physics, University of Göttingen, 37077 Göttingen, Germany}
%\affiliation{Institute for Theoretical Physics, University of Göttingen, 37077 Göttingen, Germany}
\author{Matthias Krüger}
%\affiliation{Institute for Theoretical Physics, University of Göttingen, 37077 Göttingen, Germany}
\address{Institute for Theoretical Physics, University of Göttingen, 37077 Göttingen, Germany}

\maketitle

\section{Introduction}
The analysis of particle trajectories is a fundamental building block of modern statistical physics, both in simulation and experiment. Especially in the latter, nowadays advanced techniques allow particle tracking with high precision and time resolution, leading to an extensive use of statistical analysis and the determination of microrheological and non-equilibrium properties of complex media \cite{squires_fluid_2010,krapf_power_2018,martin_comparison_2001,ahmed_active_2018,hubicka_time-dependent_2020}. Correlation functions obtained from these trajectories, like the mean squared displacement (MSD), are in equilibrium  related to powerful theorems like the fluctuation dissipation theorem (FDT) or Green-Kubo relation  \cite{kubo_fluctuation-dissipation_1966,agarwal_fluctuation-dissipation_1972,harada_equality_2005,speck_restoring_2006, baiesi_fluctuations_2009,baiesi_update_2013}. 
Breakage of time reversal symmetry is quantified by entropy production, for which powerful theorems exist \cite{zia_probability_2007,seifert_stochastic_2012,barato_thermodynamic_2015,horowitz_thermodynamic_2020, dieball_direct_2023,ghanta_fluctuation_2017,thapa_nonequilibrium_2024, knotz_entropy_2024}. 
Multi-point correlation functions, that depend on multiple time points, have the potential to be sensitive to time reversal breakage and can therefore serve as intriguing observables in non-equilibrium systems.  Conditioning trajectories, for example by selecting trajectories that start at certain positions \cite{roldan_decision_2015,berner_oscillating_2018,mori_time_2022} has lead to interesting insights. In the following, we will analyze a multi-point observable that has been termed mean back relaxation (MBR) \cite{muenker_accessing_2024, ronceray_two_2023}. It was shown that MBR is a marker for broken detailed  balance in confinement \cite{muenker_accessing_2024}. In this manuscript we expand on the  properties of this quantity and analyze it in more depth. We also extend it towards microscopic densities, which yields a marker for broken detailed balance in confinement and  for bulk systems.  

The manuscript is organized as follows. In section \ref{sec:MBR-def}, we define MBR. We   discuss its properties in confined systems in section \ref{sec:MBR-conf}, where we reproduce the proof that the deviation of the longtime value of MBR from $\frac{1}{2}$ marks broken detailed balance in confinement and a fluctuation theorem. We probe these findings on an active Brownian particle trapped in a potential. 
In section \ref{sec:PathIntegrals} we derive conditioned correlations in a path integral formalism. In section \ref{sec:MBR-Gaussian}, we  apply this to find a general relation between MBR and mean squared displacement in Gaussian systems. At last in section \ref{sec:circ-MBR} we introduce a new version of  MBR based on densities, showing that its deviation from $\frac{1}{2}$ marks  broken detailed balance in bulk systems and confinement.

\section{Mean Back Relaxation: Conditioned correlations}
\label{sec:MBR-def}

The mean back relaxation (MBR)~\cite{muenker_accessing_2024} is the average future displacement of a particle divided by the displacement in the past. For particle position coordinate $x$, we define it in terms of two time periods $t$ and $\tau$ and a cut off length $l$ as,
\begin{align}
\MBR(\tau,t,\cut) &= \mean{-\frac{x_t - x_0}{x_0 - x_{-\tau}} \reg( \left\vert x_0 - x_{-\tau} \right\vert)}\\        
     &= \int \dif x_t \dif x_0 \dif x_{-\tau} \left( - \frac{x_t - x_0}{x_0 - x_{-\tau}} \right) \times \label{eq:MBR-def} \\
    &~~~\times  \reg( \left\vert x_0 - x_{-\tau} \right\vert) W_3(x_t,t;x_0,0;x_{-\tau} , -\tau).
    \notag
\end{align}
 $W_3$ is the three point probability function, i.e., the probability of finding the particle at the positions $x_t,x_0,x_{-\tau}$ at the corresponding times. MBR is thus the ratio of the displacement $b = x_t - x_0$ in future, i.e., between $0$ and $t$ and the displacement in the past, $d = x_0 - x_{-\tau}$, i.e., between times $-\tau$ and 0.   One observes that the particle typically travels in the opposite direction or "relaxes back" compared to its earlier movement, in which case  $b$ and $d$ have opposite signs \cite{muenker_accessing_2024}. We therefore include by convention a minus sign, so that MBR is positive in these cases.

 In contrast to the definition provided in Ref.~\cite{muenker_accessing_2024}, we have made here explicit that $|d|<l$ is excluded, with cut off length $l>0$. We use in Eq.~\eqref{eq:MBR-def} a Heaviside step function $\theta(x)$ in the following factor
\begin{align}
    \reg(\left\vert x_0 - x_{-\tau} \right\vert) = \frac{\theta(\left\vert x_0 - x_{-\tau} \right\vert - l)}{\mean{\theta(\left\vert x_0 - x_{-\tau} \right\vert - l)}}.\label{eq:theta}
\end{align}
 As mentioned, the factor 
  $\reg(\vert x_\tau - x_0\vert)$ excludes the region of vanishing denominator in Eq.~\eqref{eq:MBR-def}. The denominator in Eq.~\eqref{eq:theta} is introduced for normalization \footnote{$\mean{\theta(\left\vert x_0 - x_{-\tau} \right\vert - l)} $ is nonzero if displacements $|d|\geq l$ occur.}. This definition of MBR in Eq.~\eqref{eq:MBR-def} has a  practical implementation in data from experiments or simulations; instances with $\vert x_\tau - x_0\vert < l$ are excluded from  trajectories. In other words, the object $\mean{\theta(\left\vert x_0 - x_{-\tau} \right\vert - l)}$ needs not to be evaluated explicitly. A sample code for evaluation of MBR of a Brownian particle in a harmonic potential can be found at SI. 
  In Ref.~\cite{muenker_accessing_2024}, MBR was evaluated in the limits of small $\tau$ and small $\cut$. The existence of these limits cannot be expected a priori. We continue here by analyzing MBR for finite $l$ and finite $\tau$. 
  
  To illustrate the meaning of the condition, we give the following equivalent formulation
\begin{align}
    \begin{split}
    \MBR(\tau,t,\cut) &= \int \dif d \dif b \left( -\frac{b}{d} \right) \reg(\vert d \vert) p(b,d) \\
    &= \int \dif d \left( - \frac{\mean{b}_{\vert d}}{d} \right)  \reg(\vert d \vert) p(d).    
        \end{split}\label{eq:mBR-def2}
\end{align}
Here $\mean{b}_{\vert d} = \mean{x_t - x_0}_{\vert d} = \int \dif b~[b~ p(b\vert d)]$ is the conditional average of the displacement $b = x_t - x_0$ given $d$. Eq.~\eqref{eq:mBR-def2} shows that  MBR probes how conditions affect the movement of the particle. Division by $d$, despite the need of a cut off length $l$, is advantageous, as it renders MBR dimensionless.  This  gives MBR a  quantitative meaning,  which we expect to allow useful quantitative comparison of different systems.

\section{MBR in stationarity}
\label{sec:MBR-conf}
%\subsection{Stationarity}
\subsection{Long time limit of MBR and time reversal}
In this subsection, we reproduce for completeness some of the results given in Ref.~\cite{muenker_accessing_2024}. We will in this subsection discuss stationary systems in confinement. We thus assume that a stationary distribution $W_1(x)$ with a finite mean $\mean{x}$ exists. We also assume that the joint three-point distribution function can be separated for $t\to\infty$, i.e.,\begin{subequations}
\label{eq:fact}
\begin{align}    
    \mean{x} &= \int \dif x~x~ W_1(x) ~\text{with}~  \mean{x}~\text{finite}, \label{eq:fact1} \\
   \lim_{t \to \infty}& W_3(x_t,t;x_0,0;x_{-\tau},-\tau) \label{eq:fact2}
   \\ &= W_1(x_t) W_2(x_0,0;x_{-\tau},-\tau).\notag   
\end{align}
\end{subequations}
We introduced the joint two-point probability $W_2$ \cite{risken_fokker-planck_1996}.  
In this subsection, we assume the following symmetry of $W_2$
\begin{align}
W_2(x_t,t;x_0,0) = W_2(x_0,t;x_t,0),\label{eq:sym}
\end{align}
which is the usual condition for detailed balance for the joint distribution \cite{risken_fokker-planck_1996,de_groot_non-equilibrium_1962}.
 We  insert Eq.~\eqref{eq:fact} in the definition of MBR, Eq.~\eqref{eq:MBR-def}, and obtain
\begin{align}   
    \lim_{t \to \infty} \MBR(\tau,t,\cut) &= \int \dif x_t \dif x_0 \dif x_{-\tau} \left( - \frac{x_t - x_0}{x_0 - x_{-\tau}} \right) \times \\ ~~~\times& \reg ( \vert x_0 - x_{-\tau} \vert) W_1(x_t)  W_2(x_0,0;x_{-\tau},-\tau)\notag \\      
    &= \int \dif x_0 \dif x_{-\tau} \left( - \frac{\mean{x} - x_0}{x_0 - x_{-\tau}} \right) \times \label{eq:MBR-lt} \\ &~~~\times \reg (\vert x_0 - x_{-\tau} \vert)W_2(x_0,0;x_{-\tau},-\tau) .\notag    
\end{align}
Evaluating half of Eq.~\eqref{eq:MBR-lt} by a change of variables, $x_0\leftrightarrow x_{-\tau}$ yields a form in terms of the difference of $W_2$ with arguments exchanged,
\begin{align}      
    \lim_{t \to \infty} &\MBR(\tau,t,\cut) = \frac{1}{2} +\notag\\
    &\frac{1}{2} \int \dif x_0 \dif x_{-\tau} \left(- \frac{\mean{x} - x_0 }{x_0 - x_{-\tau}} \right)    \reg (\vert x_0 - x_{-\tau} \vert) \times \notag\\
    &~~~\times \left[W_2(x_0,0;x_{-\tau}, -\tau) - W_2(x_{-\tau},0;x_0, -\tau) \right] ,\label{eq:W2s}    
\end{align}
The term in the second line vanishes when inserting Eq.~\eqref{eq:sym}. The long time value of MBR is thus
\begin{align}
    \lim_{t \to \infty} \MBR(\tau,t,\cut) \overset{\eqref{eq:fact}\&\eqref{eq:sym}}{=} \frac{1}{2}.\label{eq:MBR12}
\end{align}
We conclude that, if Eq.~\eqref{eq:MBR12} is violated and Eq.~\eqref{eq:fact} holds, Eq.~\eqref{eq:sym}, i.e., detailed balance, must be broken. Notably, the result in Eq.~\eqref{eq:MBR12} is independent of the time period $\tau$, and it is independent of the cut off length $l$. 

Eq.~\eqref{eq:MBR12} is not restricted to particle position $x$. It is generally valid for observables A with $\pi A = \epsilon A, \epsilon \in \left\lbrace -1,1 \right\rbrace$ with $\pi$ the kinematic reversal operator and the corresponding detailed balance condition $W_2(A_t,t;A_0,0) = W_2(\epsilon A_0,t; \epsilon A_t,0)$. %Thus it is valid for time  symmetric observables  (e.g., position or density) ($\epsilon = 1)$ and for time antisymmetric ones  such as velocity $(\epsilon = -1)$. 
We provide the corresponding general derivations in appendix \ref{sec:ap-Hamilton}. 

\subsection{Fluctuation theorem}
In this subsection we assume Eq.~\eqref{eq:fact} to be given, without assuming symmetry of $W_2$. 
We introduce path probability $p[x(t)]$, so that, with detailed balance, $p[x(t)] = p[\theta x(t)]$ for any path, with bracket $[\dots]$ indicating the functional dependence on $x(t)$. The  path probabilities of $x(t)$ and $\theta x(t)$ yield a fluctuation theorem for MBR. The stochastic total change of entropy for the stationary process is defined as \cite{Maes2003,seifert_entropy_2005,seifert_stochastic_2012,roldan_decision_2015}
\begin{align}
    s\left[x(t) \right] = \log \frac{p[x(t)]}{p[\theta x(t)]}\label{eq:entropy-def}.
\end{align}
Stochastic total change of entropy therefore is log ratio of the probability of a given path and the probability of its reversed. We rewrite Eq.~\eqref{eq:W2s} by change of integration variables
\begin{align}      
    \lim_{t \to \infty} &\MBR(\tau,t,\cut) \notag\\   
    &= \frac{1}{2} + \frac{1}{2} \mean{\frac{x_0 +  x_{-\tau} -2 \mean{x} }{x_0 - x_{-\tau}} \reg(\vert x_0 - x_{-\tau} \vert)}.\label{eq:needed}  
\end{align}
Eq.~\eqref{eq:needed} is reformulated in the fluctuation theorem \cite{knotz_entropy_2024}
\begin{align}
    \mean{ \left(- \frac{\mean{x} - x_0}{x_0 - x_{-\tau}} \right)\reg(\vert x_0 - x_{-\tau} \vert) \left(1 + e^{-s} \right)} = 1.
    \label{eq:MBR-FT}
\end{align}
Note that for any observable $O(x(t))$ the average of the time reversed observable can be rewritten as
\begin{align}
    \mean{O[\theta x(t)]} &= \int \Dif x(t) O[x(t)] \frac{p[\theta x(t)]
}{p[x(t)]} p[x(t)]\\ &= \mean{O[x(t)] e^{-s}}
\end{align}
using entropy production as defined in Eq.~\eqref{eq:entropy-def} and $\Dif x(t)$ denoting a path integral. The fluctuation theorem Eq.~\eqref{eq:MBR-FT} therefore states that the long-time value of MBR evaluated from  original and reversed dynamics add up to unity
\begin{align}
\begin{split}
\lim_{t \to \infty}& \left( \MBR(\tau,t,\cut) + \MBR_r(\tau,t,\cut)\right) \\ &= \left< - \frac{\mean{x} - x_0}{x_0 - x_{-\tau}}\reg(\vert x_0 - x_{-\tau} \vert) \right. \\ &\left. - \frac{\mean{x} - x_{-\tau}}{x_{-\tau} - x_{0}} \reg(\vert x_{-\tau} - x_{0} \vert)\right> = 1,\label{eq:1}
\end{split}
\end{align}
where we introduced $\MBR_r(\tau,t,\cut)$, the MBR evaluated from time backward trajectories.  Notably, the result in Eq.~\eqref{eq:1} is independent of the time period $\tau$, and it is independent of the cut off length $l$. 
In equilibrium, original and reversed dynamics are statistically indistinguishable, which is equivalent to $s = 0$ for all paths, hence the long-time value of MBR equals $\frac{1}{2}$ as stated previously.  

To sum up, in a system with Eq.~\eqref{eq:fact} fulfilled, Eq.~\eqref{eq:MBR-FT} holds, and the long-time values of MBR evaluated from original and reversed dynamics 
add up to unity, Eq.~\eqref{eq:1}. With Eqs.~\eqref{eq:fact} and  \eqref{eq:sym} fulfilled, the long-time value of  MBR equals $\frac{1}{2}$. If Eq.~\eqref{eq:fact} is fulfilled, a deviation of the long time value from $\frac{1}{2}$ thus marks the invalidity of Eq.~\eqref{eq:sym}, i.e., breakage of detailed balance.  These results are independent of time period $\tau$  and cut off length $l$.
\subsection{Example: Active Brownian particle}
\label{subsec:ABP}
To demonstrate the results of the previous subsection,  we probe a two-dimensional active Brownian particle confined in a harmonic trap of stiffness $k$, described by the Langevin equations\cite{bechinger_active_2016,dauchot_dynamics_2019}
\begin{subequations}
\label{eq:ABP}
\begin{align}
    \dot {\mathbf{x}} &= -\frac{k}{\gamma} \mathbf{x} + v_0 \mathbf{\hat{e}}(\varphi) + \mathbf{f}\\
    \dot \varphi &= f_\varphi
\end{align}
\end{subequations}
with $\mathbf{\hat{e}} = (\cos(\varphi),\sin(\varphi))^\T$ and white noise random forces $\mean{f^{(i)}(t)f^{(j)}(t^\prime)} = 2D \delta(t-t^\prime) \delta_{ij}$ and $\mean{f_\varphi (t) f_\varphi (t^\prime)} = 2D_\varphi \delta(t-t^\prime)$. $D$ and $D_\varphi$ are the translational and rotational diffusion coefficients, respectively. Simulation results are shown in Figure \ref{fig:ABP-FT}, where we use one of the coordinates to evaluate MBR. Simulation details can be found in appendix \ref{sec:ap-simulations}. If the active propulsion $v_0 = 0$ vanishes, the system is in equilibrium. The blue line in Figure \ref{fig:ABP-FT} shows this case, for $\tau=2.0 \frac{\gamma}{k}$ and $l=0.2 \sqrt{\frac{\gamma D}{k}}$,  and indeed MBR approaches the equilibrium $\frac{1}{2}$ value for large times $t$. The red lines in Figure \ref{fig:ABP-FT} are for finite $v_0$, using the same values of $\tau$ and $l$ as for the passive case. In this case the long-time MBR deviates from $\frac{1}{2}$. This result indicates the breakage of detailed balance.
Evaluation of MBR for the reversed trajectories results in a deviation from $\frac 1 2$ of opposite sign for $t\to\infty$, i.e.,  they add up to one as  stated by the fluctuation theorem Eq.~\eqref{eq:MBR-FT}. The active Brownian particle therefore demonstrates that a deviation of MBR from $\frac{1}{2}$ detects the breakage of time reversal symmetry, and this breakage is accurately described by the corresponding fluctuation theorem.

Figure \ref{fig:ABP-LT-tau} shows the long-time value of MBR as a function of $\tau$. For $v_0=0$, this longtime value is indeed independent of $\tau$, as stated by Eq.~\eqref{eq:MBR12}. 
 For finite $v_0$, i.e., with detailed balance broken, the long time value does depend on $\tau$. For $\tau\to \infty$, the long time MBR seems to approach $\frac{1}{2}$, and the deviation from $\frac 1 2$ is most pronounced for an intermediate value of $\tau$, denoted $\tau_\text{max}$. The model in Eq.~\eqref{eq:ABP} encodes three time scales, $\frac{1}{D_\varphi}$, $\frac{D}{v_0^2}$ and $\frac{\gamma}{k}$, resembling the time scales of rotational diffusion, the time scale where active and passive motion are comparable, and the time scale of relaxation in the harmonic potential, respectively.  The inset in Figure \ref{fig:ABP-LT-tau} shows the dependence of $\tau_\text{max}$ on these, by varying $\frac{1}{D_\varphi}$ and $\frac{D}{v_0^2}$ and by scaling the axes in units of $\frac{\gamma}{k}$. $\tau_{\text max}$ increases with $\frac{1}{D_\varphi}$ and with $\frac{D}{v_0^2}$, indicating  that breakage of detailed balance is sensitive to $D_\varphi$ and $v_0$:  As regards the dependence on propulsion velocity $v_0$, we interpret that for smaller $v_0$, it takes a longer time for activity to become noticeable, and hence $\tau_{\text max}$ is larger.   
  Decreasing rotational diffusion $D_\varphi$ makes the motion more persistent and thereby increases the timescales of motion. This seems to also increase the timescale $\tau_{\text max}$.

 Figure \ref{fig:ABP-LT-l} shows the long time value as a function of cut off length $l$. As expected, for $v_0=0$, this value is independent of $l$. Also, for finite $v_0$, the sum of forward and backward cases is independent of $l$. For finite $v_0$, the long time value does depend on $l$. Notably, the data suggests that the limit $l \to 0$ exists.

\begin{figure}
    \centering
    \includegraphics[width=\linewidth]{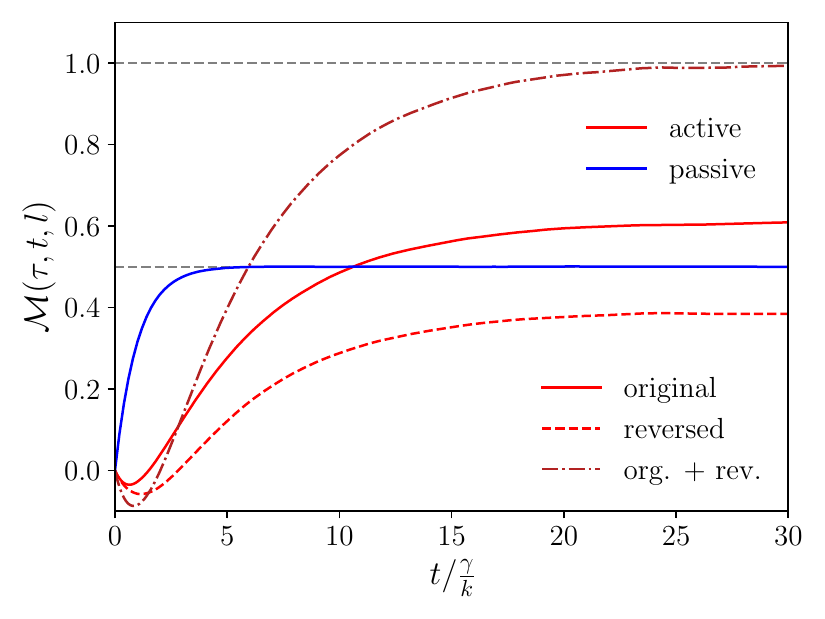}
    \caption{MBR as a function of time $t$, obtained from simulations of an active or passive Brownian particle, Eq.~\eqref{eq:ABP}, for $\tau = 2.0\frac{\gamma}{k},~\cut = 0.2 \sqrt{\frac{\gamma D}{k}}$  and $\frac{\gamma D_\varphi}{k} = 0.2$. Blue line shows a passive particle with $v_0 = 0$, i.e., a system with detailed balance, and MBR approaches the long-time value $\frac{1}{2}$. The solid red line shows an active Brownian particle with ${v_0}/ \left( {\sqrt{ \frac{k}{\gamma D}}D} \right) = 7$. The long-time value deviates from $\frac{1}{2}$, indicating a non-equilibrium system. The red dotted line is for the same system, evaluated for the time reversed trajectories. In this case, for $t\to\infty$,  MBR deviates from $\frac{1}{2}$ in opposite direction. Their sum  is shown as a dark red dotted line, which approaches unity in the long-time limit, as predicted by the fluctuation theorem Eq.~\eqref{eq:MBR-FT}. }
    \label{fig:ABP-FT}
\end{figure}

\begin{figure}
    \centering
    \includegraphics[width=\linewidth]{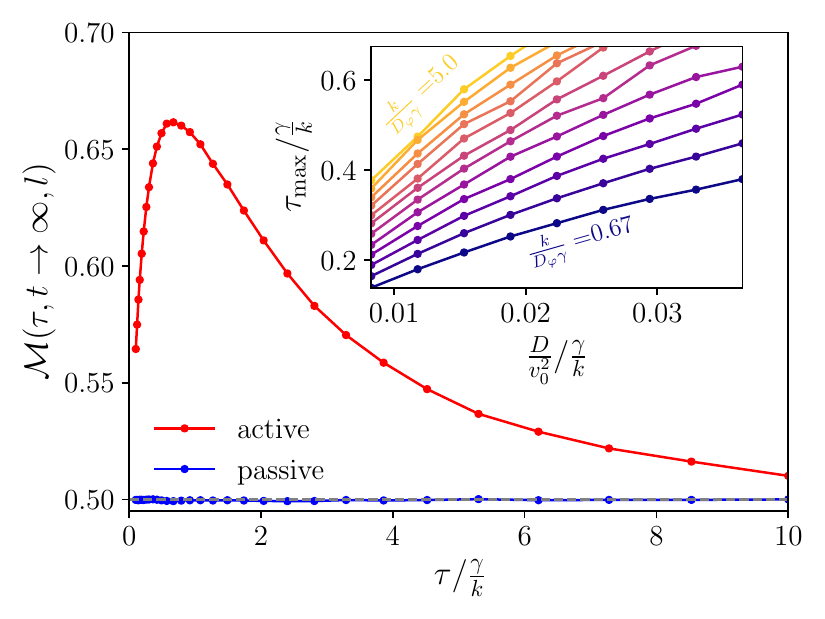}
    \caption{Long-time limit of MBR of ABP as function of $\tau$ with $l=0.2\sqrt{\frac{\gamma D}{k}}$ and $\frac{\gamma D_\varphi}{k} = 0.2$. In the passive case $v_0 = 0$ (blue line) the MBR is $\frac{1}{2}$ independent of $\tau$. In the active case with ${v_0}/ \left( {\sqrt{ \frac{k}{\gamma D}}D} \right) = 7$ (red line), MBR depends on $\tau$, and  the deviation from $\frac 1 2$ shows a maximum, denoted $\tau_{\text max}$. The inset shows  $\tau_\text{max}$ as a function of $\frac{D}{v_0^2}$, for different $\frac{1}{D_\varphi}$, with all times in units of $\frac{\gamma}{k}$. Different line colors indicate different $\frac{1}{D_\varphi}$, equidistant 
    between $\frac{k}{D_\varphi\gamma}=0.67 $ and $\frac{k}{D_\varphi\gamma}=5.0 $ (corresponding to a spacing of $\approx 0.394$ between lines).}
    \label{fig:ABP-LT-tau}
\end{figure}

\begin{figure}
    \centering
    \includegraphics[width=\linewidth]{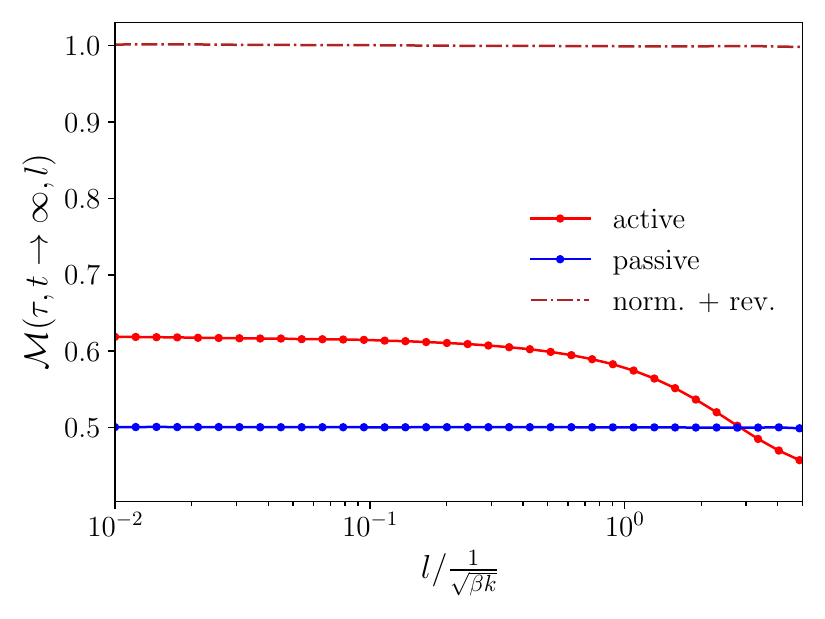}
    \caption{ The long-time limit of MBR of ABP as function  of $l$ with $\tau=2.0\frac{\gamma}{k}$ and $\frac{\gamma D_\varphi}{k} = 0.2$. In the passive case $v_0 = 0$  (blue line) there is no dependence, as expected from Eq.~\eqref{eq:MBR12}. In the active case ${v_0}/ \left( {\sqrt{ \frac{k}{\gamma D}}D} \right) = 7$ (red line) MBR depends on $l$, with the limit  $l \to 0$ apparently existing. The fluctuation theorem Eq.~\eqref{eq:MBR-FT} does not depend on $l$ shown by the red dotted line for ${v_0}/ \left( {\sqrt{ \frac{k}{\gamma D}}D} \right) = 7$.}
    \label{fig:ABP-LT-l}
\end{figure}

\newpage

\section{Conditioned correlations and MBR in Path Integrals}
\label{sec:PathIntegrals}
\subsection{Conditioned path integrals}
We provide here the derivation of  MBR in a path integral formalism. This can then directly be used to compute MBR for a Gaussian system in section \ref{sec:MBR-Gaussian}. The path integral is formally given by \cite{altland_condensed_2010}
\begin{align}
    1 = \int \Dif x(t) e^{A[x(t)]} 
\end{align}
with probability $p[x(t)] = e^{A[x(t)]}$ with the action $A[x(t)]$ which is a functional of the path $x(t)$. To compute correlations of particle positions we introduce the moment generating functional \cite{hertz_path_2016}
\begin{align}
    Z[j(t)] =  \int \Dif x(t) e^{A[x(t)] + \int \dif s j(s) x(s)}.
\end{align}
Any moment of positions is calculated by taking functional derivatives with respect to $j(t)$, which we will refer to as the source term. In case of the mean value
\begin{align}
    \mean{x(t)} = \frac{1}{Z[0]} \left. \frac{\delta Z[j]}{\delta j(t)} \right\vert_{j=0}.
\end{align}
with $\frac{\delta}{\delta j(t)}$ the functional derivative at time $t$. To condition the functional on traveling a distance $d$ in a time interval $\tau$, we include a delta-function
\begin{align}
    \hat{Z}[j] &= \int \Dif x(t) \delta(x(0) - x(-\tau) - d ) e^{A[x(t)] + \int \dif s j(s) x(s)}\\
        &= \int \Dif x(t) \frac{\dif q}{2\pi}~ e^{-iqd} \times \\
        &~~~\times e^{A[x(t)] + \int \dif s x(s)(j(s) + iq\delta(s) - iq\delta(s + \tau))} \notag\\        
        &= \int \frac{\dif q}{2\pi} ~e^{-iqd} Z[j(t) + iq\delta(t) - iq\delta(t+\tau)].
\end{align}
By using the Fourier representation of the delta function, we expressed the conditioned generating functional by a Fourier transformation of the original generating functional with an altered source term $j(t) + iq\delta(t) - iq\delta(t+\tau)$. The conditioned mean position is then given by
\begin{align}
    \mean{x(t)}_{\vert d} &= \left. \frac{1}{\hat{Z}[0]} \frac{\delta \hat{Z}[j]}{\delta j(t)} \right\vert_{j=0}\\
    &= \frac{\left. \frac{\delta}{\delta j(t)} \int \frac{\dif q}{2\pi} ~e^{-iqd} Z[j(t) + iq\delta(t) - iq\delta(t+\tau)] \right\vert_{j=0}}{\int \frac{\dif q}{2 \pi} ~e^{-iqd} Z[iq\delta(t) - iq\delta(t+\tau)]} \label{eq:cond-mean}
\end{align}
Note that this expression is equivalent to
\begin{widetext}
\begin{align}
    \mean{x(t)}_{\vert d} - \mean{x(0)}_{\vert d} &= \left. \frac{\frac{\partial}{\partial (ik)} \int \Dif x(t) \frac{\dif q}{2\pi} e^{iq(x(0) - x(-\tau) - d)} e^{ik(x(t) - x(0))} e^{A[x(t)]}}{\int \Dif x(t) \frac{\dif q}{2\pi}  e^{iq(x(0) - x(-\tau) - d)} e^{ik(x(t) - x(0))} e^{A[x(t)]}} \right\vert_{k=0}\\
    &= \left. \frac{ \frac{\partial}{\partial(ik)}\int \frac{\dif q}{2 \pi}e^{-iqd}\mean{e^{iq(x(0) - x(-\tau))} e^{ik(x(t) - x(0))}}}{\int \frac{\dif q}{2 \pi} e^{-iqd}\mean{e^{iq(x(0) - x(-\tau))} e^{ik(x(t) - x(0))}}} \right\vert_{k=0} \label{eq:cond-mean-char}
\end{align}
\end{widetext}
which technically only requires ordinary and no functional derivatives. Notably, the term $\mean{e^{iq(x(0) - x(-\tau))} e^{ik(x(t) - x(0))}}$  is a four point correlation of microscopic density $e^{ikx}$  \cite{dhont_studies_1996, hansen_theory_2013}. The MBR is explicitly calculated via 
\begin{align}
    \MBR(\tau,t,\cut) &= \mean{-\frac{x_t - x_0}{d} \reg (\vert d \vert)}\\
    &=-\int \dif d \frac{\mean{x(t)}_{\vert d} - \mean{x(0)}_{\vert d}}{d} \reg (\vert d \vert) p(d)
\end{align}
with $p(d)$ the probability of the particle displacement $d$ in time $\tau$. We omit here the explicit derivation of $p(d)$ in the path formalism.
\section{MBR for Gaussian Systems -- MBR-MSD formula}
\label{sec:MBR-Gaussian}
We will in this section find MBR for a stationary, one dimensional Gaussian process by applying Eq.~\eqref{eq:cond-mean-char}. Examples for  such processes are harmonically coupled Brownian particles, which are widely used to  model complex material behaviors \cite{doerries_correlation_2021,khan_trajectories_2014, ginot_recoil_2022,caspers_how_2023}.
\subsection{Relating MBR and MSD for Gaussian systems}
\label{subsec:Gaussian}
Consider a stationary process with displacements $x(t) - x(0)$ Gaussian distributed. For simplicity, we consider displacements with zero mean, i.e., $\mean{x(t) - x(0)} = 0$ for any $t$. % We make no other assumption about the dynamics. 
The four point correlator of Eq.~\eqref{eq:cond-mean-char} is given for this case by
\begin{multline}
     \mean{e^{iq(x(0) - x(-\tau))} e^{ik(x(t) - x(0))}} =  \\
     \exp \biggl( -\frac{1}{2} \mean{(x(0) - x(-\tau))^2} q^2 - \frac{1}{2} \mean{(x(t) - x(0))^2} k^2 \\
       - \mean{(x(t) - x(0))(x(0) - x(-\tau)) }qk \biggr). \hspace{1cm}   
 \label{eq:char-Gauss}
 \end{multline}
Inserting Eq.~\eqref{eq:char-Gauss} into Eq.~\eqref{eq:cond-mean-char} gives
\begin{align}    
    \begin{split}
     \mean{x(t)}_{\vert d} - \mean{x(0)}_{\vert d} = \hspace{4.6cm}\\
     \frac{\int \dif q (iq) \left[ \mean{(x(t) - x(0))(x(0) - x(-\tau))} \right]  e^{-iqd} e^{ -\frac{q^2}{2\sigma^2}}}{\int \dif q e^{-iqd} \exp \left( -\frac{q^2}{2\sigma^2} \right)}     
     \end{split}
\end{align}
with $\frac{1}{\sigma^2} = \mean{(x(0) - x(-\tau))^2}$.  After completing the square, the calculation reduces to evaluation of the first moment of a Gaussian distribution in $q$ with mean $-id \sigma^2$ and variance $\sigma^2$. The final result for the conditioned mean is given by
\begin{multline}
    \mean{x(t)}_{\vert d} - \mean{x(0)}_{\vert d} =\\  d \frac{  \mean{(x(t) - x(0))(x(0) - x(-\tau))}}{\mean{(x(0) - x(-\tau))^2}}.
\end{multline}
Notably, the conditioned mean is linear in $d$, which renders MBR independent of cutoff length $l$. From this expression we can evaluate  MBR via
\begin{align}
    \MBR(\tau,t,\cut) &= -\int \dif d \frac{\mean{x(t)}_{\vert d} - \mean{x(0)}_{\vert d}}{d}\reg (\vert d \vert) p(d)\\
    &= -\frac{  \mean{(x(t) - x(0))(x(0) - x(-\tau))}}{\mean{(x(0) - x(-\tau))^2}}.
    \label{eq:MBR-tmp}
\end{align}
Note that integration over the probability distribution $p(d)$ is trivial after division by $d$, as it then reduces to unity by normalization, i.e., $1 = \int \dif d~ \reg (\vert d \vert) p(d)$.
A useful reformulation of Eq.~\eqref{eq:MBR-tmp} is obtained by expressing the correlations in terms of the mean squared displacement (MSD) $\Delta x^2(t) = \mean{(x(t) - x(0))^2} = 2\mean{x^2} - 2\mean{x(t)x(0)}$. This transformation relates the MBR in Gaussian systems with the MSD,
\begin{align}
    \MBR(\tau,t,\cut) &= \frac{1}{2} \left(1 - \frac{\Delta x^2(t+\tau) - \Delta x^2(t)}{\Delta x^2(\tau)} \right).
    \label{eq:MBR-msd}
\end{align}
The MBR-MSD formula, \eqref{eq:MBR-msd}, is the main result of this section. As MBR in \eqref{eq:MBR-msd} is independent of the cut off $l$, we will  omit this argument in the remainder of this section for brevity. 
We will discuss conclusions for special cases in the following subsections.

In is insightful to recall that a stationary Gaussian process of zero mean is completely characterized by its mean squared displacement. It is thus no surprise that MBR in this case can be expressed in terms of MSD.

\subsection{Example: Overdamped particles}
\label{subsec:TwoCoupled}
\begin{figure*}
    \centering    
    \includegraphics{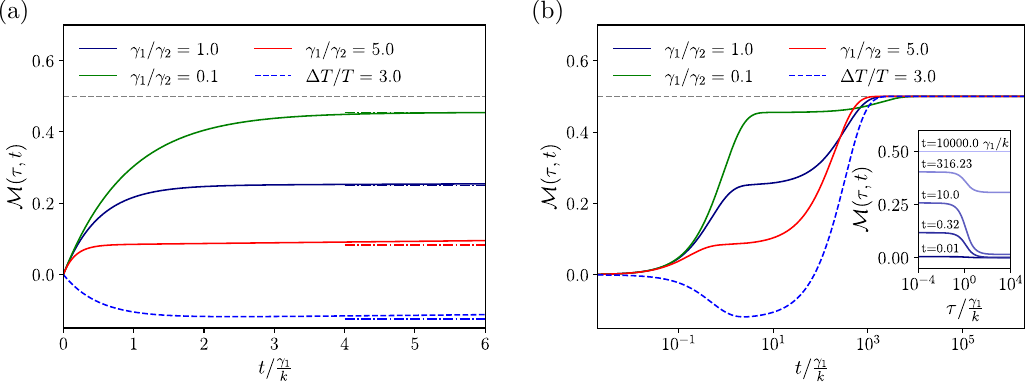}
\caption{MBR for two coupled Brownian particles in a harmonic potential with $\frac{k^\prime}{k} = 0.003$ (Eq.~\eqref{eq:Two-part}, analytical solution using Eq.~\eqref{eq:limtau0}). a) shows the small time window $t \ll \tau_2$. Solid lines give equilibrium cases $T_1 = T_2 = T$ for different ratios of friction coefficients $\gamma_i$. Dotted line shows $\frac{\Delta T}{T} = 3$ and $\frac{\gamma_1}{\gamma_2} = 1$. Horizontal lines at the right indicate the plateau of Eq.~\eqref{eq:plateau}. (b) Same curves on logarithmic time axis, illustrating the different time regimes, including approach of $\frac{1}{2}$ for $t\gg\tau_2$. Inset shows the dependence of the MBR on $\tau$ at different $t$ for $\frac{\gamma_1}{\gamma_2}= 1$. }
    \label{fig:TwoParticles}
\end{figure*}

As an example, consider two  Brownian particles coupled by a harmonic spring. We also include an external trapping potential, so that the system is stationary,
\begin{subequations}
\label{eq:Two-part}
\begin{align}
    \dot x_1 &= -\frac{k^\prime}{\gamma_1} x_1 - \frac{k}{\gamma_1} (x_1 - x_2) + f_1(t)\\
    \dot x_2 &= -\frac{k^\prime}{\gamma_2} x_2 - \frac{k}{\gamma_2} (x_2 - x_1) + f_2(t),
\end{align}
\end{subequations}
with coupling constants $k$ and  $k^\prime$, friction coefficients $\gamma_i$ and random forces $\mean{f_i(t) f_j(t^\prime)}= \frac{2 k_B T_i}{\gamma_i} \delta_{ij}$. The particles may have different temperatures $T_1 = T, T_2 = T + \Delta T$, so that cases out of equilibrium can be considered.
The MSD of $x_1$ is for this linear system found analytically, it reads in the considered  stationary state, (see details in appendix \ref{sec:ap-TwoCoupled})
\begin{align}
\begin{split}
\Delta x_1^2(t) =   \frac{2k_B T}{k} 
    &\left[ \Psi(\tau_1,\tau_2) (1 - e^{-t/\tau_1}) \right.\\
    +&\left.~  \Psi(\tau_2,\tau_1) (1 - e^{- t/\tau_2}) \right]\label{eq:MSD}    
    \end{split}
\end{align}
with the timescales $\tau_{1,2}$ and the dimensionless function 
$\Psi$ given in appendix \ref{sec:ap-TwoCoupled}. MBR of the particle is then found via Eq.~\eqref{eq:MBR-msd}. We restrict, for simplicity of discussion, to the limit of small $\tau$. The MSD in Eq.~\eqref{eq:MSD} is diffusive for small times, and Eq.~\eqref{eq:MBR-msd} can be rewritten in the limit of $\tau \to 0$,
\begin{align}
    \lim_{\tau \to 0} \MBR(t,\tau) = \frac{1}{2} \left( 1 - \frac{\left. \frac{\dif}{\dif t^\prime}  \Delta x^2(t^\prime) \right\vert_{t^\prime = t}}{\left. \frac{\dif}{\dif t^\prime}  \Delta x^2(t^\prime) \right\vert_{t^\prime = 0}}\right).\label{eq:limtau0}
\end{align}
In this limit, MBR is thus related to time derivatives of  MSD. We will use Eq.~\eqref{eq:limtau0} in the following.

It is insightful to consider the case of $k'\ll k$, for which the time scales separate, i.e., in this limit, $\tau_1\ll\tau_2$, with
\begin{align}
    \tau_1&= \frac{\gamma_1}{k} \frac{\gamma_2}{ \gamma_1 + \gamma_2}\\
    \tau_2&=  \frac{\gamma_1 + \gamma_2}{2  k^\prime}.
\end{align}
In this limit, $\tau_1$ is the relaxation time of the bond between the particles, and $\tau_2$ is the time of relaxation in the external trap. 

For $t\ll\tau_1$, the particle dynamics is that of a free particle, i.e.,  $\Delta x^2(t)= 2 t k_BT/\gamma_1$, and MBR in Eq.~\eqref{eq:limtau0} vanishes. A free Brownian particle shows zero MBR.

For $\tau_1\ll t\ll\tau_2$, MSD shows an intermediate diffusive regime,
\begin{align}
\Delta x^2(t)= 2 t \left[\frac{k_B T}{\gamma_1 + \gamma_2} + \frac{k_B \Delta T \gamma_2}{(\gamma_1 + \gamma_2)^2}\right].
\end{align}
In this time window, $\MBR$, using Eq.~\eqref{eq:limtau0}, is nearly time independent, i.e., it shows a plateau of height%Eq.~\eqref{eq:largetsmalltau}
\begin{align}
    \frac{1}{2} \left(1 - \frac{\gamma_1}{\gamma_1 + \gamma_2} - \frac{\Delta T}{T} \frac{\gamma_1 \gamma_2}{(\gamma_1 + \gamma_2)^2} \right) .  \label{eq:plateau}
\end{align}
Fig.~\ref{fig:TwoParticles} shows example curves of MBR of this model, for a ratio of $k'/k=0.003$, in which the  two time scales are well separated.

Fig.~\ref{fig:TwoParticles}a) shows the time window  $t\ll\tau_2$, demonstrating that the curves indeed plateau at values given by Eq.~\eqref{eq:plateau}. Notably, for $\Delta T=0$, the plateau of Eq.~\eqref{eq:plateau} is bound between 0 and $\frac{1}{2}$, and it can take negative values for finite $\Delta T$. 

For $t\gg\tau_2$, the MSD  saturates to its final, time independent value. In this limit, MBR found via Eq.~\eqref{eq:limtau0}, approaches the value  $\frac{1}{2}$, for any values of parameters in Eq.~\eqref{eq:Two-part}.   Fig.~\ref{fig:TwoParticles}b) uses a logarithmic time axis to illustrate the three mentioned time regimes, displaying the final approach to $\frac{1}{2}$ of all curves shown.  

It is important to realize that the stochastic Gaussian 1d process of $x_1$ in Eq.~\eqref{eq:Two-part} is  time reversal symmetric \cite{netz2023multipoint, muenker_accessing_2024}, i.e.,  Eq.~\eqref{eq:sym} holds, for any sets of parameters. This is the reason why, MBR, for $t\gg \tau_2$ must approach $\frac{1}{2}$ to obey  Eq.~\eqref{eq:MBR12}.

We may use this example to illustrate the case of "bulk" referred to in Ref.~\cite{muenker_accessing_2024}. Let $\tau_2$ be so large (i.e., $k'$ be so small), that MBR can only be resolved (e.g. due to experimental limitations) for $t\ll \tau_2$. MBR curves in this case may look similar as shown in Fig.~\ref{fig:TwoParticles}a) \cite{muenker_accessing_2024}. One has to keep in mind that, during this time window, where MSD still grows as a function of time, a deviation of MBR from $\frac{1}{2}$ is not related to broken detailed balance. This observation was stated as "MBR is not a marker for broken detailed balance in bulk" in Ref.~\cite{muenker_accessing_2024}.   

We conclude this section by  noting that the model of Eq.~\eqref{eq:Two-part} may be extended to an arbitrary number of connected overdamped degrees. In a setting of equilibrium, i.e., all temperatures equal, the corresponding  Smoluchwoski equation \cite{dhont_studies_1996} allows an 
  expansion in eigenfunctions to yield for the MSD \cite{risken_fokker-planck_1996}
 of one of the degrees
\begin{align}
    \Delta x^2(t) = \sum_n \left\vert c_n \right\vert^2 \left( 1 -e^{-\lambda_n t} \right) .\label{eq:cor-smol}
\end{align}
The eigenvalues $\lambda_n$ are real and non-negative, as are the expansion coefficients $\left\vert c_n \right\vert^2$. Inserting Eq.~\eqref{eq:cor-smol} into Eq.~\eqref{eq:MBR-msd}, we find
\begin{align}
    \MBR(\tau,t) = \frac{1}{2} \left(1 - \frac{\sum_n \left\vert c_n \right\vert^2 (1 - e^{-\lambda_n \tau} ) e^{-\lambda_n t}}{\sum_n \left\vert c_n \right\vert^2 (1 - e^{-\lambda_n \tau} )} \right).
    \label{eq:MBR-smol}
\end{align}
From Eq.~\eqref{eq:MBR-smol}, MBR for this overdamped equilibrium system, is a monotonic function of time, and rests between $0$ and $\frac{1}{2}$
\begin{align}
0&\leq \frac{d}{dt} \MBR(\tau,t),\\
0&\leq \MBR(\tau,t) \leq \frac{1}{2}.
    \end{align}
\section{Density Mean Back Relaxation marks broken detailed balance in confinement and bulk}
\label{sec:circ-MBR}
As emphasized in section \ref{sec:MBR-Gaussian},
for large systems (sometimes termed as bulk), it may take a long time for the limit of Eq.~\eqref{eq:fact2} to be approached, i.e., the long time limit of Eq.~\eqref{eq:MBR12} may not be reachable in practice. %for detecting  broken detailed balance challenging.
 
This challenge can be overcome by introducing observables that show well defined mean values even in the limit of large system size. We propose here to use the microscopic density in Fourier space \cite{hansen_theory_2013} (for simplicity restricting to a one dimensional wave vector $q$)
\begin{align}
    \rho_q(t) = \sum_{j=1}^N e^{iq x_j(t)}. \label{eq:struct}
\end{align}
Here, $j$ runs over the different particles of an $N$ particle system. $\rho_q$ is the Fourier transform of the microscopic density in position space, $\rho(x)=\sum_{j=1}^N \delta(x-x_j)$ \cite{hansen_theory_2013}.  We continue under the assumption that the mean  $\langle \rho_q \rangle$ is unique and well defined. This assumption rests on the observation that the mean density $\langle \rho(x)\rangle$ is finite for typical particle interactions \cite{hansen_theory_2013}, so that
\begin{align}
    \mean{\rho_q} = \int \dif x \mean{\rho (x)} e^{iqx}
\end{align}
is finite as well. For example, for homogeneous systems, the mean density is constant in space $\mean{\rho(x)} = \rho_0$, and $\mean{\rho_q} = \rho_0 \delta(q)$ vanishes for any finite $q$ \cite{hansen_theory_2013}. 

\begin{figure}
    \centering    
    \includegraphics[width=\linewidth]{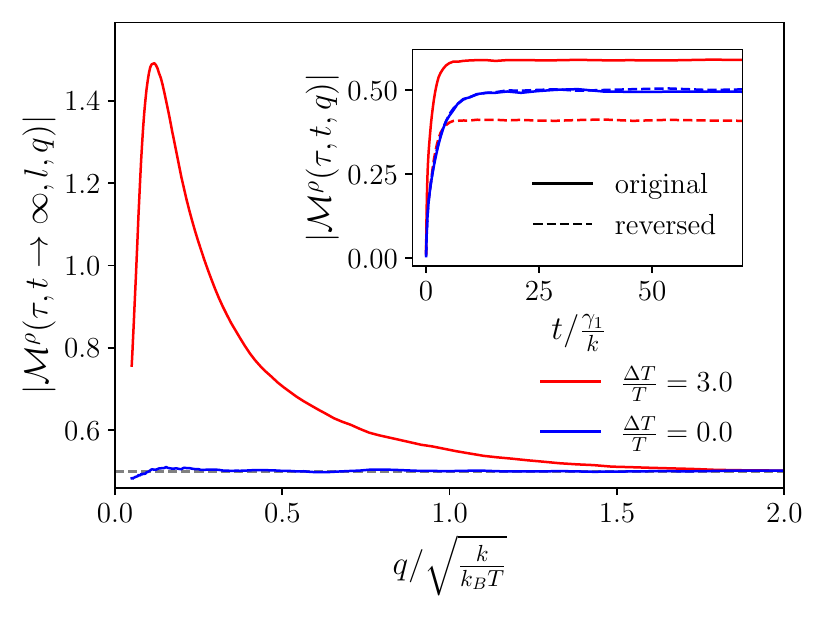}
    \caption{Density MBR, Eq.~\eqref{eq:dMBR-def}, for the model of two coupled particles of  Eq.~\eqref{eq:Two-part}, with $\tau=0.1 \frac{\gamma_1}{k}$, $\frac{\gamma_1}{\gamma_2} = 1, \frac{k^\prime}{k} = 0.003$ and $l = 0.01$ ($l$ is dimensionless for dMBR). Main plot shows the long time limit as a function of $q$, displaying the value of $\frac{1}{2}$ for equal temperatures, and a deviation for different temperatures.  The deviation from $\frac{1}{2}$ is maximal for $q_\text{max} \approx 0.11 \sqrt{ \frac{k}{k_B T}}$ for this set of parameters.    
    Inset shows the behavior for $q=0.8 \sqrt{ \frac{k}{k_B T}}$ as a function of time, where we note that the numerical data obey Eq.~\eqref{eq:dMBR-1}.    
  }
    \label{fig:dMBR}
\end{figure}

We thus introduce the  {\it density} mean back relaxation dMBR,
\begin{align}
    \MBR^\rho(\tau,t,\cutq,q) = \mean{- \frac{\rho_q(t) - \rho_q(0)}{\rho_q(0) - \rho_q(-\tau)} \regq (\vert \rho_q(0) - \rho_q(-\tau) \vert)}. \label{eq:dMBR-def}
\end{align}
 dMBR  additionally depends on the wave vector $q$ compared to MBR.
 A global displacement  $x_i +a $ yields a factor $\rho_q(t) \to  e^{iq a} \rho_q(t)$, which cancels in Eq.~\eqref{eq:dMBR-def}.  dMBR does thus not depend on the choice of coordinate origin. Most important,  we can thus state
\begin{align}
    \lim_{t\to\infty}\MBR^\rho(\tau,t,\cutq,q) \overset{\eqref{eq:fact-A} \& \eqref{eq:db-A}}{=}\frac{1}{2}, \label{eq:dMBR-12}
\end{align}
so that a deviation of dMBR from $\frac{1}{2}$ marks broken detailed balance. Furthermore, the sum of long time values of dMBR for original and reversed trajectories adds up to unity, analogously to Eq.~\eqref{eq:1},
\begin{align}
\lim_{t\to\infty}\MBR^\rho(\tau,t,\cutq,q)+ \lim_{t\to\infty}\MBR^\rho_r(\tau,t,\cutq,q) =1. \label{eq:dMBR-1}
\end{align}
To demonstrate these properties, we revisit the two coupled particles of Eq.~\eqref{eq:Two-part}, see Fig.~\ref{fig:dMBR}. 
Indeed, for $T_1=T_2$, dMBR approaches $\frac{1}{2}$, while it deviates from $\frac{1}{2}$ for $T_1\not=T_2$, with Eq.~\eqref{eq:dMBR-1} fulfilled. As expected, the deviation from $\frac{1}{2}$ depends on $q$.  We expect that the dependence on $q$ yields information on the   length scales of the system.

\section{Discussion}
\label{sec:Discussion}
We discussed several properties of the recently introduced mean back relaxation, and extended the definition by including a cut off length $l$, so that MBR depends, additionaly to observation time $t$, on the parameters of time period $\tau$ and length $l$. 
In stationary confined systems, deviation of the long time value of  MBR from $\frac{1}{2}$ marks breakage of detailed balance \cite{muenker_accessing_2024}. With or without detailed balance, MBR from forward and reversed paths sum up to unity. These statements are true for any values of $\tau$ and $l$. We exemplify this for the case of a trapped active Brownian particle. 

We point out that  Eq.~\eqref{eq:MBR12} is generally valid   for  observables that are even or odd under time reversal. Future work will investigate MBR with particle velocity, which is odd under time reversal.

Using a path integral approach we find a general solution for  MBR in terms of multipoint density correlations. Evaluating this for Gaussian systems yields a relation between mean squared displacement (MSD) and MBR, from which additional properties of MBR in Gaussian systems  can be inferred. In Gaussian overdamped equilibrium systems, MBR is monotonic in time, and takes values between $0$ and $\frac{1}{2}$. We exemplified the relation between MBR and MSD via the example model of two trapped coupled Brownian particles. Studying the regime of parameters for which the relaxation times are well separated allows for additional insights: For intermediate times, the MSD is diffusive, and MBR takes (nearly) time independent values, which show a strong dependence on the chosen parameters. For even longer times, MSD approaches a constant, and MBR approaches $\frac{1}{2}$ for any set of parameters. We point out that Eq.~\eqref{eq:MBR12} is valid for the latter time regime only, and the value of $\frac{1}{2}$ is found because 1d stationary Gaussian systems obey detailed balance.

We define  a density MBR, dMBR, using the Fourier transform of microscopic density as observable. dMBR enjoys the same properties as MBR: It marks breakage of detailed balance, and sums up to unity from normal and reversed paths. We illustrate its advantages using the model of coupled Brownian particles: i) The final value of MBR is typically approached  faster compared to position MBR; its relaxation time remains finite when the trapping potential vanishes, in contrast to that of position MBR. ii) Density MBR uses the positions of both particles as an input, and it detects broken detailed balance for the case of different temperatures, in contrast to position MBR. 

Future work will investigate how the relaxation time of density MBR can be estimated a priori.

 It  remains open whether monotonicity and positivity of MBR in equilibrium  can be proven more generally. For dMBR, more examples need to be studied, by testing it in many-body systems
\cite{hub_short-range_2007,grimm_simple_2022}.

\section*{Acknowledgements}
We thank Timo Betz, Till Moritz Muenker,  Tim Salditt and Dmitrii E. Makarov
for fruitful discussions.
\appendix

\section{Eq.~\eqref{eq:MBR12} for general observable}
\label{sec:ap-Hamilton}
We state the assumptions of Eqs.~\eqref{eq:fact} and \eqref{eq:sym} for a general phase space observable $A$,
\begin{subequations}
\label{eq:fact-A}
\begin{align}
    \mean{A} &= \int \dif A~ A  W_1(A) ~\text{with }\mean{A}~\text{finite}\\
    \lim_{t \to \infty} W_3&(A_t,t;A_0,0;A_{-\tau},-\tau)\\
    &= W_1(A_t) W_2(A_0,0;A_{-\tau},-\tau) \notag
\end{align}
\end{subequations}
We also repeat the symmetry condition for $W_2$ \cite{de_groot_non-equilibrium_1962}
\begin{align}
    W_2(A_t,t;A_0,0) = W_2(\epsilon A_0,t; \epsilon A_t,0),
    \label{eq:db-A}
\end{align}
denoted as the usual condition for detailed balance for the joint distribution in Ref.~\cite{risken_fokker-planck_1996}. We assume that $A$ is either symmetric ($\epsilon=1$) or antisymmetric ($\epsilon=-1$) under path reversal.

Assuming Eq.~\eqref{eq:db-A} and stationarity, one can use that $W_1(A) = \int \dif A_t W_2(A_t,t;A,0) = \int \dif A_0 W_2(A,t;A_0,0)$ to find 
\begin{align}
\label{eq:db-mean}
\begin{split}
    \mean{A} &= \int \dif A_t \dif A_0~ A_0 W_2(A_t,t;A_0,0)\\    
    & =  \int \dif A_t \dif A_0~ \epsilon A_t W_2(A_t,t;A_0,0)\\
    &= \epsilon \mean{A}.        
\end{split}
\end{align}
We proceed as in the main text, 
we start by the definition of the MBR and insert Eq.~\eqref{eq:fact-A} for the long time limit
\begin{align}
    \lim_{t \to \infty} 2 \MBR^A&(\tau,t,l) \overset{\eqref{eq:fact-A}}{=} \int \dif A_0 \dif A_{-\tau} \left(- \frac{\mean{A} - A_0}{A_0 - A_{-\tau}} \right) \notag \\
    &\times \reg (\vert A_0 - A_{-\tau} \vert) W_2(A_0,0;A_{-\tau},-\tau) \\
    &+ \int \dif A_0 \dif A_{-\tau} \left(- \frac{\mean{A} - \epsilon A_{-\tau}}{\epsilon A_{-\tau} - \epsilon A_{0}} \right) \notag\\
    &\times \reg (\vert \epsilon A_0 - \epsilon A_{-\tau} \vert) W_2(\epsilon A_{-\tau},0; \epsilon A_0,-\tau). \notag
\end{align}
We used the coordinate transform $A_0 \to \epsilon A_{-\tau}$ and $A_{-\tau} \to \epsilon A_0$ in the second summand. By adding a zero in the counter of the second summand and algebraic transformations one gets, using Eq.~\eqref{eq:db-mean}
\begin{align}
    \lim_{t \to \infty} &\MBR^A(\tau,t,l) = \frac{1}{2} \int \dif A_0 \dif A_{-\tau} \frac{A_0 - A_{-\tau}}{A_0 - A_{-\tau}} \\
    &\times \reg (\vert  A_0 -  A_{-\tau} \vert) W_2(\epsilon A_{-\tau},0; \epsilon A_0,-\tau) \notag\\
    &+\frac{1}{2} \int \dif A_0 A_{-\tau} \left( - \frac{\mean
{A} - A_0}{A_0 - A_{-\tau}} \right)
\reg (\vert  A_0 -  A_{-\tau} \vert) \notag\\
&\times   \left[ W_2(A_0,0;A_{-\tau},-\tau)- W_2(\epsilon A_{-\tau},0; \epsilon A_0,-\tau) \right] \notag
\end{align}
The first term is evaluated as $\frac{1}{2}$ because of the normalization $\mean{\reg (\vert  A_0 -  A_{-\tau} \vert)} = 1$. Using Eq.~\eqref{eq:db-A} we find that the second term vanishes and obtain
\begin{align}
    \lim_{t \to \infty} \MBR^A(\tau,t,l) & \overset{\eqref{eq:fact-A} \& \eqref{eq:db-A}}{=} \frac{1}{2}.
\end{align}
\section{Simulations}
\label{sec:ap-simulations}
Simulations are performed with the Julia "Differential Equations" package with a Runge-Kutta-Milstein method (RKMil) without adaptive time stepping. Internal time step is $dt = 10^{-3}$ and the particle position is saved at $\Delta t = 10^{-2}$ in units of the simulation time scale. The simulations are performed in dimensionless space and time. For the active Brownian particle length is given in units of $\sqrt{\frac{\gamma D}{ k}}$ and time in units of  $ \frac{\gamma}{k}$. Therefore the following dimensionless SDEs were simulated (compare Eq.~\eqref{eq:ABP})
\begin{subequations}
\label{eq:ABP_dimensionless}
\begin{align}
    \dot {\mathbf{x}} &= - \mathbf{x} + \frac{v_0}{\sqrt{\frac{k}{\gamma D} }D} \mathbf{\hat{e}}(\varphi) +\mathbf{f}\\
    \dot \varphi &= f_\varphi
\end{align}
\end{subequations}
with $\mean{f^{(i)}(t)f^{(j)}(t^\prime)} = 2 \delta(t-t^\prime) \delta_{ij}$ and $\mean{f_\varphi (t) f_\varphi (t^\prime)} = 2\frac{D_\varphi \gamma}{ k} \delta(t-t^\prime)$. These equations thus involve two dimensionless parameters, ${v_0}/\left({\sqrt{\frac{k}{\gamma D}  }D} \right)$ and $\frac{D_\varphi \gamma}{k}$.

For the coupled particles of Eq.~\eqref{eq:Two-part}, we scale length in terms of $ \sqrt{\frac{k_B T}{ k}}$ and time 
 in terms of  $ \frac{\gamma_1}{k}$. The dimensionless equations are (compare Eq.~\eqref{eq:Two-part})
\begin{subequations}
\label{eq:Two-part-siml}
\begin{align}
    \dot x_1 &= - \frac{k^\prime}{k} x_1 -  (x_1 - x_2) + f_1(t)\\
    \dot x_2 &= - \frac{k^\prime}{k} \frac{\gamma_1}{\gamma_2} x_2 - \frac{\gamma_1}{\gamma_2} (x_2 - x_1) + f_2(t)
\end{align}
\end{subequations}
with $\mean{f_1(t) f_1(t^\prime)} = 2 \delta(t-t^\prime)$ and $\mean{f_2(t) f_2(t^\prime)} = 2 \frac{\gamma_1}{\gamma_2} \frac{T + \Delta T}{T} \delta(t-t^\prime)$ which depend on the dimensionless parameters $\frac{\gamma_2}{\gamma_1}$, $\frac{T + \Delta T}{T}$ and $\frac{k^\prime}{k}$. 

Figures 1-3 have a trajectory length of $t = 10^5$ in units of the simulation time scale with 80 trajectories, inset Figure 2 trajectory length of $t=10^4$ with 1400 trajectories. Long time MBR in Figure 2 and 3 are evaluated at $t = 40$. Figure 5, main, has a trajectory length of $t=10^4$ and 80 trajectories, and inset trajectory length of $t=400$ and 800 trajectories.
Figure 5 long time MBR evaluated at $t = 1000$.

\section{Two coupled particles in harmonic potential}
\label{sec:ap-TwoCoupled}
The Langevin equations of the two linearly coupled particles, discussed in section \ref{subsec:TwoCoupled}, are given by
\begin{subequations}
\begin{align}
    \dot x_1 &= - \frac{k^\prime}{\gamma_1} x_1 -\frac{k}{\gamma_1} (x_1 - x_2) + f_1\\
    \dot x_2 &= - \frac{k^\prime}{\gamma_2}  -\frac{k}{\gamma_2} (x_2 - x_1) + f_2
\end{align}
\end{subequations}
which is rewritten into the matrix equation
\begin{align}
    \dot {\mathbf{x} } = -\mathbf{M} \cdot\mathbf{x} + \mathbf{f}
\end{align}
with
\begin{align}
    \mathbf{M} = \left( \begin{array}{cc}
         \frac{k^\prime + k}{\gamma_1}& -\frac{k}{\gamma_1}  \\
         -\frac{k}{ \gamma_2}& \frac{k^\prime + k}{\gamma_2}
    \end{array} \right)
\end{align}
which has the general solution
\begin{align}
    \mathbf{x}(t) = \int_{-\infty}^t e^{-\mathbf{M} (t-s)} \cdot\mathbf{f}(s).
\end{align}
To solve this equation we diagnoalize $\mathbf{M}$ with the eigenvalues
\begin{align}
\begin{split}
    \lambda_{1/2} = &\frac{(k^\prime + k)(\gamma_1 + \gamma_2)}{2\gamma_1 \gamma_2}\\
    &\pm \sqrt{\left( \frac{(k^\prime + k)(\gamma_1 + \gamma_2)}{2\gamma_1 \gamma_2} \right)^2 - \frac{{k^\prime}^2 + 2 k k^\prime}{\gamma_1 \gamma_2}}
    \end{split}
\end{align}
and corresponding eigenvectors $(1,  \frac{k^\prime}{k} + 1 - \frac{\gamma_1}{k}\lambda_i)^\intercal = (1,a(\lambda_i))^\intercal$. After basis transform one gets the following expression
\begin{align}
    x_1(t) &= \frac{1}{a(\lambda_2) - a(\lambda_1)}\\
    &\int_{-\infty}^t \left[ \left( a(\lambda_2) e^{-\lambda_1 (t-s)} - a(\lambda_1) e^{-\lambda_2 (t-s)} \right) f_1(s) \right. \notag \\ &+ \left. \left(-e^{\lambda_1 (t-s) } + e^{\lambda_2 (t-s)} \right) f_2(s) \right] \dif s  \notag.
\end{align}
From here we can evaluate the correlation function
\begin{align}
    \begin{split}    
    \mean{x_1(t) x_1(0)} = \hspace{0.6\linewidth}\\ \frac{k_B T }{k} \left[ \left( \phi(\lambda_1, \lambda_2) + \left(1 + \frac{\Delta T}{T} \right) \psi(\lambda_1, \lambda_2) \right) e^{-\lambda_1 t}\right.\\
    \left. + \left( \phi(\lambda_2, \lambda_1) + \left(1 + \frac{\Delta T}{T} \right) \psi(\lambda_2, \lambda_1) \right) e^{-\lambda_2 t}\right]
    \end{split}
\end{align}
with dimensionless
\begin{align}
    \phi(\lambda_1 , \lambda_2) &= \frac{k^3}{\gamma_1^3} \frac{1}{(\lambda_2 - \lambda_1)^2} \left( \frac{a(\lambda_2)^2}{\lambda_1}  - \frac{2a(\lambda_1)a(\lambda_2)}{\lambda_1 + \lambda_2}  \right)\\
    \psi(\lambda_1, \lambda_2) &= \frac{k^3}{\gamma_1^3} \frac{\gamma_1}{\gamma_2} \frac{1}{(\lambda_2 - \lambda_1)^2} \left( \frac{1}{\lambda_1} - \frac{2}{\lambda_1 + \lambda_2} \right)
\end{align}

The mean squared displacement is obtained by the identity
\begin{align}
    \Delta x^2(t) &= \mean{(x(t) - x(0))^2} \notag\\
    &= 2 \mean{x^2} - 2 \mean{x(t) x(0)}
\end{align}
and we get
\begin{align}
\begin{split}
    \Delta x_1^2(t) = \frac{2 k_B T}{k} \left[ \Psi(\lambda_1,\lambda_2)(1 - e^{-\lambda_1 t}) \right.\\
    \left. +\Psi(\lambda_2,\lambda_1)(1 - e^{-\lambda_2 t}) \right]        
\end{split}
\end{align}
with $\Psi(\lambda_1, \lambda_2) = \phi(\lambda_1, \lambda_2) + (1 + \frac{\Delta T}{T}) \psi(\lambda_1,\lambda_2)$. The relation is rewritten into the timescales $\tau_i = \frac{1}{\lambda_i}$. In the main text we investigate the case $k^\prime \ll k$. In this case the first time scale $\tau_1 = \frac{1}{\lambda_1}$ approaches
\begin{align}
    \lim_{k^\prime/k \to 0} \frac{1}{\lambda_1} = \frac{1}{k}\frac{\gamma_1\gamma_2}{\gamma_1 + \gamma_2}
\end{align}
However in the same limit
\begin{align}
    \lim_{k^\prime/k \to 0} \lambda_2 = 0
\end{align}
so the second time scale diverges as one would expect. It diverges with a power law
\begin{align}
   \lim_{k^\prime/k \to 0}\frac{k^\prime}{k}\tau_2  = \lim_{k^\prime/k \to 0}\frac{\frac{k^\prime}{k}}{\lambda_2} = \frac{\gamma_1 + \gamma_2}{2 k}
\end{align}
which means $\tau_2 \approx \frac{\gamma_1 + \gamma_2}{2 k^\prime}$. On time scales $\tau_1 \ll t \ll \tau_2$ we approach a diffusive regime for which we can find a diffuson coefficient by expanding in terms of $\lambda_2 t$
\begin{align}\begin{split}
    \Delta x_1^2(t) = \frac{2 k_B T}{k} \left[ \Psi(\lambda_1,\lambda_2) (1 - e^{-\lambda_1 t}) \right.\\
    + \left. \Psi(\lambda_2,\lambda_1) \lambda_2 t + \mathcal{O}(\lambda_2^2 t^2) \right]
    \end{split}
\end{align}
where we identify $D_L = \frac{k_BT}{k} \lambda_2 \Psi(\lambda_2,\lambda_1)$ as the diffusion coefficient. In the limit $\frac{k^\prime}{k} \to 0$ one gets
\begin{align}
    D \approx \frac{k_B T}{\gamma_1 + \gamma_2} + \frac{k_B \Delta T \gamma_2}{(\gamma_1 + \gamma_2)^2} 
\end{align}
\bibliographystyle{ieeetr}
\bibliography{references}
\end{document}